# Revive, Restore, Revitalize:
# An Eco-economic Methodology for Maasai Mara

He Sun[1], Junfeng Zhu[2], Yipeng Xu[3]

University of Nottingham Ningbo China

## Abstract

Kenya's Maasai Mara is known as a popular wildlife preserve and a significant habitat for many species. However, extensive human activities have led to the endangerment of some species and the degradation of ecosystems. To improve the management of resources both inside and outside the reserve, we propose a dynamic system that balances the ecosystem and human interests.

For the model establishment and preparation part, we built an **agent-based model** consisting of 71 animal species, 10 human categories, and 2 natural resource types that simulate the savanna ecosystem in Maasai Mara to some extent. The model uses **metabolic rate-mass relation** to calculate the energy consumed and gained by animals, logistic curves to simulate the growth of animals, individual interactions to simulate the whole food web among wildlife, and the impact of human intervention. **Fitness proportional selection** and **particle swarm algorithm** have been introduced to simulate the preferences of various organisms for different natural resources.

To specify and quantify the outcomes of preserve-related actions, we developed 21 management policies related to tourism, transportation, taxation, environmental pollution, animal research, diplomacy, and poaching varied by areas using a game-theoretic approach. We weighed four development indicators - environment, research, economy, and security - using the **TOPSIS method**. Each of these policies affects one or more among 16 factors (which generate the 4 indicators) with various weights to some extent, thereby resulting in different influences on multiple aspects of the model. We evaluate these effects in order to reduce negative interactions between animals and humans and to balance the interests of people.

We ranked these policies and grouped their effects on different aspects into three clusters: Environmental Paradise, Economic Thrive, and Comprehensive Development. These policy clusters are applied to the simulated animal-human-nature interaction in the ecosystem dynamics. We then use the **entropy weight method** to evaluate the Preserve Development Index of these policy and strategy clusters over a ten-year period to determine the most effective cluster that considers both environmental and economic factors.

Our model makes predictions for ten years after policy implementation, from which we can witness that policies with comprehensive development characteristics have more potential. In terms of long-term results, our model is robust and adaptive. If our model were to be transferred to other wildlife refuges, this could be done by altering the modeling of the maps, the attributes and the densities of animals and humans, and the distribution of natural resources.

[1] scyhs4@nottingham.edu.cn
[2] scyjz14@nottingham.edu.cn
[3] ssyyx20@nottingham.edu.cn

# Contents

# 1 Introduction

## 1.1 Problem Background

"We must work together to engage communities to see wildlife's value, not just for themselves but for the world's future generations," U.S. Ambassador to Kenya Michael Godike spoke on World Wildlife Day on March 3, 2016[1].

Kenya's wildlife preserves have long been recognized as crucial areas for protecting natural resources and preserving wildlife. However, as human populations have expanded and the demands for economic growth have increased, the management of these preserves has become more complex. To address these challenges, the Kenyan government has passed the Wildlife Conservation and Management Act, 2013, which seeks to promote more equitable resource sharing and community-based management efforts[2].

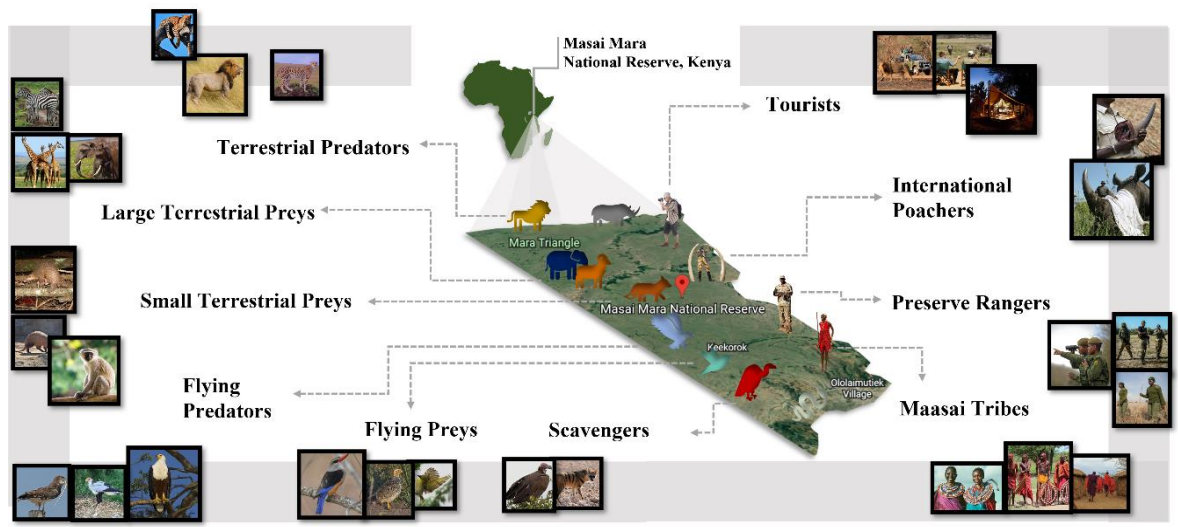

**Figure 1: Animals and Humans in the Maasai Mara Preserve and Nearby Areas**

One particular preserve, the Maasai Mara, presents unique management challenges. The preserve is home to a diverse array of wildlife, including many endangered species, and is a major tourist attraction for the region. However, the area is also home to many local communities who rely on the land for their livelihoods. The challenge, therefore, is to develop policies and management strategies that protect the wildlife and natural resources within the preserve while also balancing the needs and interests of the local people.

As a result, there is a need for a methodology to determine which policies and management strategies will result in the best outcomes. This methodology should take into account the potential economic impacts of different management approaches, as well as the likelihood of negative interactions between animals and humans. Ultimately, the goal is to develop a plan that will ensure the long-term sustainability of the preserve and its surrounding areas, while also promoting economic growth and development for the local communities.

Overall, the proposed plan has significant value for the Maasai Mara and other wildlife management areas. By taking a comprehensive approach that balances the interests of both wildlife and local communities, it is possible to create a sustainable management strategy that benefits everyone involved.

## 1.2 Restatement of the Problem

Considering the analysis of the background information and the restricted conditions of the problem statement, the following problems need to be specifically solved:

- **Develop specific policies and management strategies** for different areas within the Maasai Mara Preserve that protect wildlife and natural resources while **balancing the interests** of local communities and mitigating the negative impacts on people and animals.

- Create a methodology to **determine the best policies and management strategies** for the Maasai Mara Preserve. This methodology should rank and compare outcomes and predict the interactions between animals and people, as well as the resulting economic effects in the area within and around the preserve.

- Provide predictions about the long-term trends resulting from the **specified policies and strategies** and analyze the possible **long-term outcomes**. Additionally, describe how this approach could be applied to other wildlife management areas. Finally, provide a non-technical report discussing the proposed plan and its value for the Kenyan Tourism and Wildlife Committee.

## 1.3 Modeling Framework

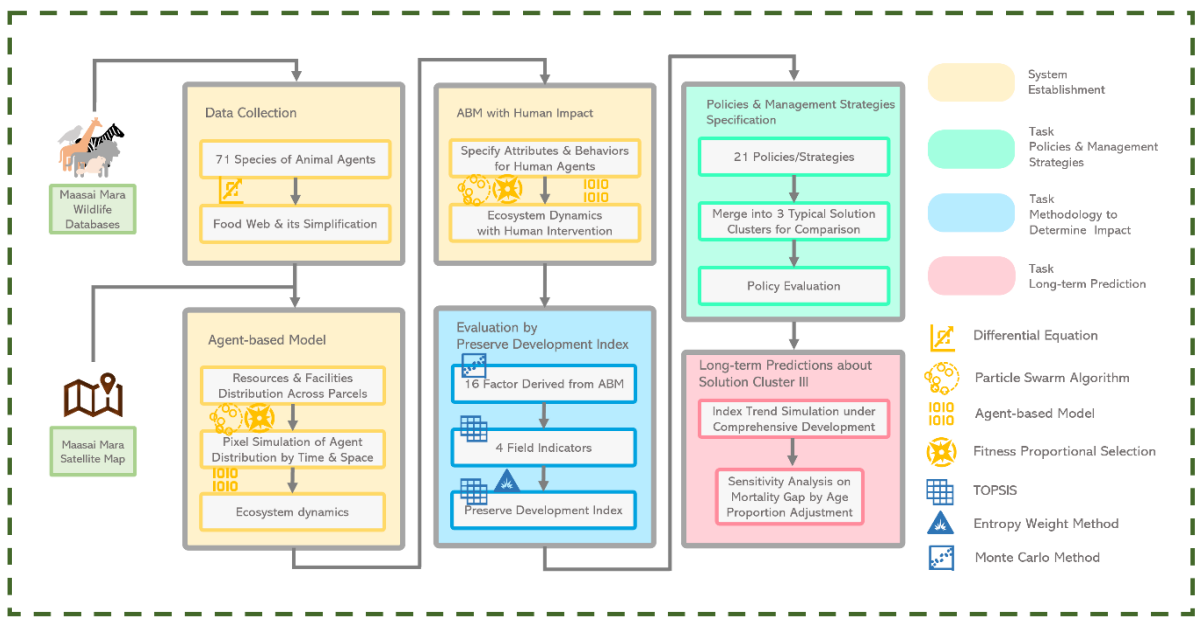

Figure 2. The Mind Map of Our Work

- Collect information on the **distribution of animals**, types of **humans** and **resources** in the Maasai Mara Preserve.

- Developing an Agent-based Model (ABM) for a stable **ecosystem without humans**.

- Involving humans in ABM to describing and predicting **human-animal interactions** in ecosystems.

- Developing an **evaluation system** to derive a **Preserve Development Index** to measure the status of the Maasai Mara Preserve.
- Specifying Policies and management strategies related to different aspects for parcels in **different areas** of the preserve, for areas **inside and outside the preserve,** and for **different types of humans.**
- The different policies and management strategies are **classified into different clusters** according to the **aspects of their effects**, and the different clusters are compared in terms of their effects. Finally, the **best clusters of policies and strategies** are identified.
- Predicting the **future trends** of the Maasai Mara Preserve under different situations.
- Finally, a sensitivity analysis of the model is performed to verify its **robustness and reliability**. And analyze the feasibility of its application to other wildlife management fields.

## 2 Assumptions and Justifications

- **During the time scale we discuss in this paper, the earth's ecological environment keeps stable.**
  In our model, we don't consider the incidents that may cause great climate changes such as meteorite impacts or frequent volcanoes, since their probabilities are small.
- **The same agents of animals in the Maasai Mara reserve have the same birth rate.**
  This is specified based on the reproductive cycles of the different animals recorded in the official database of the Maasai Mara Conservancy, since for a large number of animals over a long-time span, the birth rate of animals has a uniform rate of increase.
- **Life history traits are stationary for agents of each species of animal or vary in response to human activity.**
  Life history traits, including growth rate, age at maturity and reproductive output, are stationary. Because the effects of factors such as genetic variation, phenotypic plasticity and trade-offs between life history traits on population dynamics vary over a time span that is usually measured in centuries[3].
- **All deaths of animal agents other than predation and human interaction are classified as random mortality.**
  Random mortality includes, but is not limited to, animal mortality due to accident, disease and environmental stochasticity, as the probability of a random mortality event occurring in a large number of individual animals is stable.
- **Interactions between animal agents are local, i.e., animals are thought to interact primarily with other animals in their vicinity.**
  Because the majority of terrestrial animals have barely global knowledge of the systems in which they live[4]. But local interactions can still produce emergent phenomena on a global scale.

# 3 Notations

The key mathematical notations used in this paper are listed in Table 1.

**Table 1: Notations Used in this Paper**

| Symbol | Description | Unit |
|--------|-------------|------|
| $t$ | Independent variable: time | month |
| $w$ | Weight of an animal agent | kilogram |
| $n$ | The total amount of an animal agent | / |
| $r$ | Growth rate in terms of animal agents' weight | percentage |
| $k$ | Energy provided by each prey as food in animal agents | colonies |
| $e$ | The hunting effectiveness of animal agents | percentage |
| $g$ | The age of animal agents | year |
| $c^*$ | Average monthly energy consumption factor of animal agent | percentage |
| $m^*$ | The average number of offspring that an individual animal agent can reproduce each month | / |
| $h^*$ | Natural resource endowment factors for different parcels | / |
| $p$ | The probability of an agent moving in a certain direction | percentage |
| $v$ | Coordinate-based moving tendency vectors | / |
| $l$ | Horizontal distance at the geographical level | kilometer |
| $\alpha$ | Environmental Indicator | / |
| $\beta$ | Scientific Research Indicator | / |
| $\gamma$ | Economic Indicator | / |
| $\delta$ | Safety Indicator | / |
| $\zeta$ | Simpson's Diversity Factor | / |
| $\eta$ | Total Biomass Density Factor | / |
| $\theta$ | Habitat Integrity Factor | / |
| $\iota$ | Animal Population Stability Factor | / |
| $\kappa$ | Zoological Research Factor | / |
| $\lambda$ | Anthropological Research Factor | / |
| $\mu$ | Citizens Income Factor | / |
| $\nu$ | Maasai Tribesmen Income Factor | / |
| $\xi$ | Local Employment Factor | / |
| $\sigma$ | Field Safety Factor | / |
| $\tau$ | Assembly Occupancy Security Factor | / |
| $\upsilon$ | Animal Protection Factor | / |
| $\phi$ | Anti-poaching Factor | / |
| $\chi$ | Animal Medication Factor | / |
| $\rho$ | Organic Resource Integrity Factor | / |
| $\psi$ | Water Integrity Factor | / |
| $\Omega$ | Preserve Development Index | / |

# 4 Agent-based Model (ABM) for the Maasai Mara Preserve

## 4.1 Data Description

Data about **animals, resources and facilities distribution** in the Maasai Mara Preserve and surrounding areas have been collected from authoritative databases. These data are used for **modelling of animal agents** and **parcel divisions.**

Table 2: Data Source Collation

| Data Names | Data Source Names | Link |
|---|---|---|
| Bird distribution | BirdLife International | http://datazone.birdlife.org/site/factsheet/masai-mara-iba-kenya |
| The distribution of some mammals | Maasai Mara National Park | https://www.maasaimara-kenyapark.com/information/animals/ |
| Elephant distribution | African Elephant Database | africanelephantdatabase.org |
| Facilities distribution | Tripadvisor | https://www.tripadvisor.com/ |
| Facilities distribution | Ecotourism Kenya | https://ecotourismkenya.org/eco-rated-facilities/eco-rated-facilities-gold-rated/entry/158/ |
| Animal size and weight | San Diego Humane Society | https://resources.sdhumane.org/Programs_and_Services/Adoption_Resources/Animal_Size_and_Weight_Chart |
| Animal specifications | Stack Exchange Public Database | https://biology.stackexchange.com/questions/79964/public-database-of-animal-specifications |
| Animal weights | Wolfram Data Repository | https://datarepository.wolframcloud.com/resources/Sample-Data-Animal-Weights |
| Animal weights and metabolic rate | Nature: Scientific Data | https://www.nature.com/articles/s41597-022-01364-9 / |
| Animal lifespan | Norwegian Veterinary Institute | https://norecopa.no/3r-guide/anage-database-of-animal-ageing-and-longevity/ |
| Animal aging and lifespan | Healthcare Data Library | https://www.johnsnowlabs.com/marketplace/the-animal-aging-and-longevity-database/ |
| Animal aging and lifespan | data.world | https://data.world/animals/zoo-animal-lifespans |

Due to the extraordinary diversity of animals in the Masai Mara reserve, a total of 71 typical species were selected as animal agents. The 71 animal agents cover six broad categories: terrestrial predator, large terrestrial prey, small terrestrial prey, flying predator, terrestrial scavenger, flying prey. The animals in these six broad categories form a relatively complete food web covering both land and air. Together with organic resources including plants as well as water resources, they form the ABM model for the Maasai Mara Preserve.

Based on a selection of **71 typical animal species** as animal agents and their **four attributes**, combined with the **interaction behaviors** between animals and the **area in which they are located**, an **agent-based model (ABM)** of the Masai Mara Preserve can be developed.

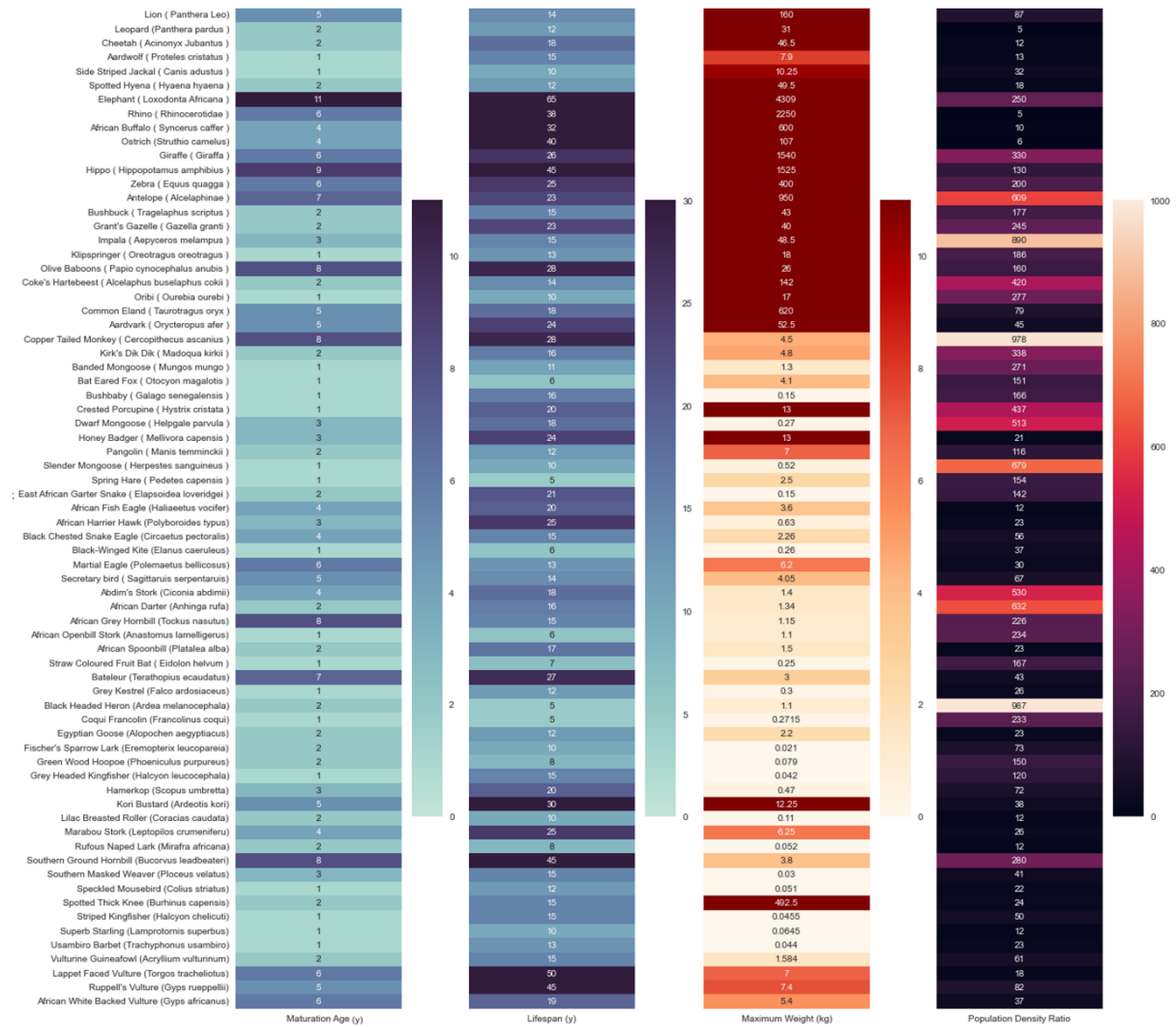

**Figure 3: 71 Species of Animal Agents Selected and Their Four Attributes**

The four attributes of the 71 animal agents are maturation age, lifespan, maximum weight and population density ratio. Animal agents cannot have offspring until they reach maturation age. If an animal agent does not die from lack of food or predation, then the maximum life span of each animal agent is its corresponding life span, and if an animal agent has enough time, it can gradually increase in weight until it reaches maximum weight. Based on the real situation in the Maasai Mara Preserve, the initial population distribution of each animal agent in the preserve is given by its population density ratio.

## 4.2 ABM Simulating Animals in Specific Environments

To use ABM to model this system, the first step is to **define the agents and their characteristics.** In this section, agents are wild animals in a protected area. Each agent has **unique attributes**, including body weight, growth rate, energy expenditure, age at maturity, predation relationships and movement speed. In addition, the animals also **interact with the environment** in a number of ways, e.g., reaching for water sources to drink.

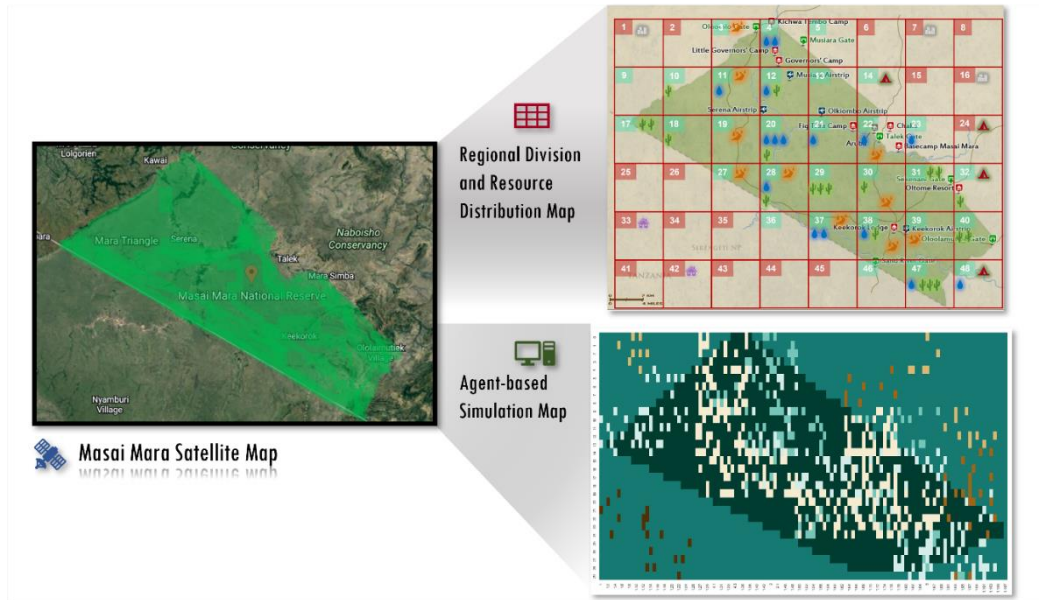

**Figure 4: Simulation of Animal Populations and Parcel Divisions in ABM.**

Based on **Metabolic Rate-Mass Relation**. For predatory-prey relationships of animal agents, the change in weight of individual animals **after each predation** is $dw/dt$, which is given by:

$$\begin{cases} w(0) = w_0 \\ \dfrac{dw}{dt} = r(t)\, w(t)\, (1 - \dfrac{w(t)}{w_{\max}}) \end{cases} \quad (1)$$

where the $w_0$ is the average weight of a specific animal, the $r(t)$ in the above function is defined as **growth rate** as a function of time in terms of animal agents' weight, $r(t)$ is given by:

$$r(t) = r_{pt}\, \dfrac{f(t)}{m(t)} \quad (2)$$

where the $r_{pt}$ is the optimal growth rate collected from the authoritative animal database. In the function (2), $m(t)$ represents the amount of food that an animal agent **needs to consume** per unit of time, $f(t)$ represents the amount of prey that a single animal agent **can obtain** in a single predation.

The derivation of $m(t)$ is as follows:

$$m(t) = a\, (w(t))^b \quad (3)$$

where $a$, $b$ are the **constant parameters** proposed in the work of Nagy, Girard, and Brown[5], with values of 139065 for $a$ and 0.889 for $b$.

The derivation of $f(t)$ is as follows:

$$f(t) = k\, e(t)\, p(t)\, \dfrac{w(t)}{w_{\max}} \quad (4)$$

where $k$ is energy provided by each prey, $p(t)$ represents the **number of preys** located within one kilometer of a particular animal agent.

The **hunting effectiveness of animal agents**, $e(t)$ in relation to age, $g$ is derived as follows:

$$\begin{cases} e(t) = e_0 + \dfrac{e_{max} - e_0}{g_{matured}} y & (g < g_{matured}) \\ e(t) = e_{max} & (g > g_{matured}) \end{cases} \quad (5)$$

Animal agents also consume energy for their daily activities, and the resulting weight loss is associated with the **energy consumption factor, $c^*$**.

The **state transfer equation** describing the **month-to-month weight change** of an animal agent is derived as follows:

$$w(t) = c^* w(t-1) \quad (6)$$

The weight of the animal agent is **limited by the maximum weight**. If it has always had sufficient food, it will continue to gain weight until it reaches its maximum weight. Conversely, if the weight i**s below a certain lower limit**, the agent will die. Alternatively, if the animal agent **is preyed upon**, it will die.

For the reproduction of animal agents, we specify that only animals that have reached the age of sexual maturity can reproduce, so that a function of the total number of animals, $n(t)$ can be derived from the number of existing mature animals, $n_{matured}$ by the reproduction factor, $m^*$.

$$n(t) = n_{matured} e^{(m^* t)} - n_{dead} \quad (7)$$

where the $n_{dead}$ is the number of dead animals due to starvation, accidents, predation, etc.

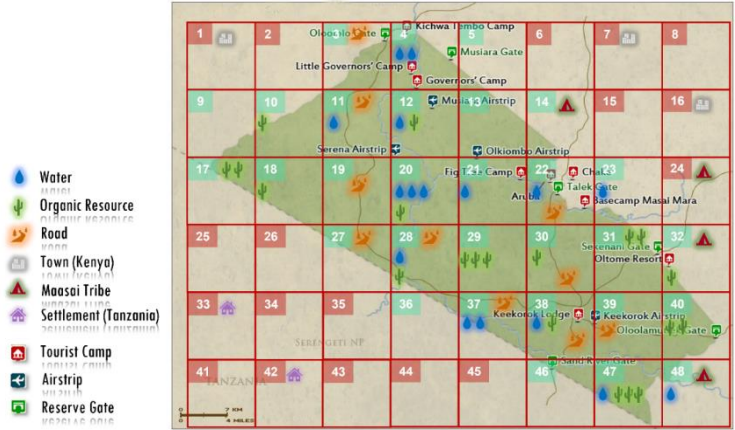

**Figure 5: The Division of Parcels and the Distribution of Resources and Facilities in Each Parcel.**
In addition, considering the **resource natural endowment of the different areas** of the preserve, herbivores and birds often **gravitate towards water and organic resource**[6]. Based on the concept of **Particle Swarm Algorithm (PSA)**, representing the attractiveness of water and organic resource to individuals acting as animal agents as vectors, the following vector function can be obtained.

$$\begin{cases} \vec{v}_{water}(x, y) = \sum h_{(x,y)} \vec{v}_{water(x,y)} \\ \vec{v}_{organic}(x, y) = \sum h_{(x,y)} \vec{v}_{organic(x,y)} \end{cases} \quad (8)$$

where the $h_{(x,y)}$ is the natural resource endowment factors for a parcel that are located within one kilometer of an individual animal agent. The $\vec{v}_{(x,y)}$ is a coordinate-based moving tendency vector generated by a parcel with one of the natural resource endowments. Considering that March to May is the long rainy season and November and December the short rainy season[7], the natural resource endowment factor, $h^*$ **becomes 1.2 times higher** in the above months than at other times.

A weighted sum of the two moving tendency vectors gives:

$$\vec{v}_{total} = a\vec{v}_{water} + b\vec{v}_{organic} \tag{9}$$

where the *a* is the **weight of an animal agent's preference** for a water and *b* is the weight of their preference for organic resource. Preference weights, *a* and *b* are both habits of a particularly animal agent[8].

The coordinates of the moving tendency vectors can be derived.

$$(x_p, y_p) = \vec{v}_{total} - (x_0, y_0) \tag{10}$$

where the $(x_0, y_0)$ is the coordinates of the location of the animal agent.

Based on the concepts of **Fitness Proportional Selection (FPS)**, the probability of an agent moving in a certain direction, the probability of an animal agent moving, *p* in a certain direction can be derived:

$$p_{w/n/e/s} = \frac{l_{e/s/w/n}}{l_e + l_s + l_w + l_n} \tag{11}$$

where $p_{w/n/e/s}$ is the probability that the animal agent moves in the four directions of west, north, east and south in once movement. $l_{e/s/w/n}$ is the distance of the coordinates of the moving tendency vectors from the reserve boundary in the east, south, west and north directions respectively. This makes animal agents **more tendentious to move closer to water and** organic resource**.**

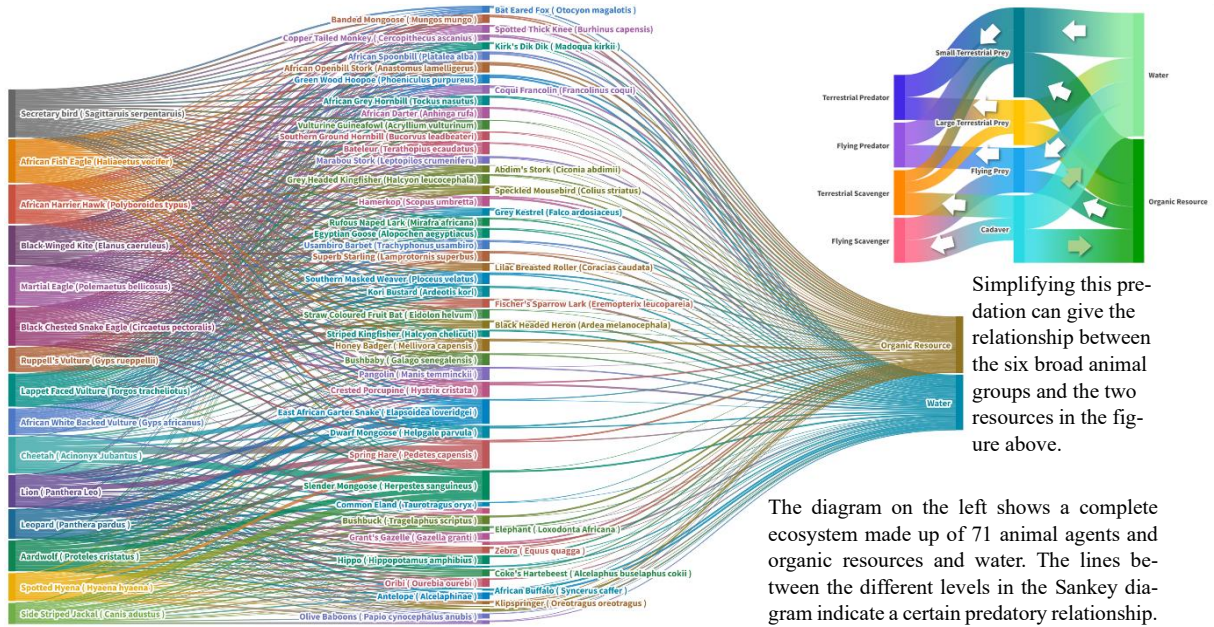

Simplifying this predation can give the relationship between the six broad animal groups and the two resources in the figure above.

The diagram on the left shows a complete ecosystem made up of 71 animal agents and organic resources and water. The lines between the different levels in the Sankey diagram indicate a certain predatory relationship.

**Figure 6. Composition of Food web**

In the food web, the predatory relationship between animal agents consists of a terrestrial predator, large terrestrial prey, small terrestrial prey, flying predator, terrestrial scavenger, flying prey. The terrestrial predator preys on large terrestrial prey and small terrestrial prey, and the flying predator preys on small terrestrial prey and flying prey.

Based on these mechanisms, an agent-based model of the animals and environment in the Maasai Mara Preserve can be developed. In this stage of the ABM, the animal evolution of the preserve in the absence of humans is simulated. While verifying the self-sustainability of this ABM, the results of the simulations will also be used for comparison with the situation after the human has been imported in the following section.

## 4.3 Time and Spatial Distribution of Animal Agents

After specifying the **initial parameters i.e. the population density ratios at the beginning of the simulation** to the ABM, the simulation results of the ABM are visualized. We can obtain the distribution of the three types of animal agents at **1, 5 and 20 years** after the start of the simulation, respectively. Considering the clarity of the ABM visualization, we have selected **several typical animals** from each of the three types of animal agents for visualization.

**Table 3: Population Density Ratio of Animal Agents**

| Animal Agents \ Time Point | Initial | 1st Year | 5th Year | 20th Year |
|---|---|---|---|---|
| Lion (Panthera Leo) | 24 | 22 | 22 | 18 |
| Antelope (Alcelaphinae) | 85 | 83 | 70 | 68 |
| Bat Eared Fox (Otocyon Magalotis) | 30 | 28 | 28 | 27 |
| Martial Eagle (Polemaetus Mellicosus) | 150 | 142 | 106 | 116 |
| Egyptian Goose (Alopochen Aegyptiacus) | 40 | 32 | 50 | 34 |
| Spotted Hyena (Hyaena Hyaena) | 16 | 16 | 15 | 8 |
| Ruppell's Vulture (Gyps Rueppellii) | 17 | 16 | 15 | 7 |

Table 3 shows the population density ratios of the selected seven species of animals at the beginning of the simulation, in the 1st year, in the 5th year and in the 20th year. The seven animals (in order from top to bottom in Animal Agents) represent terrestrial predator, large terrestrial prey, small terrestrial prey, flying predator, flying prey, terrestrial scavenger and flying scavenger.

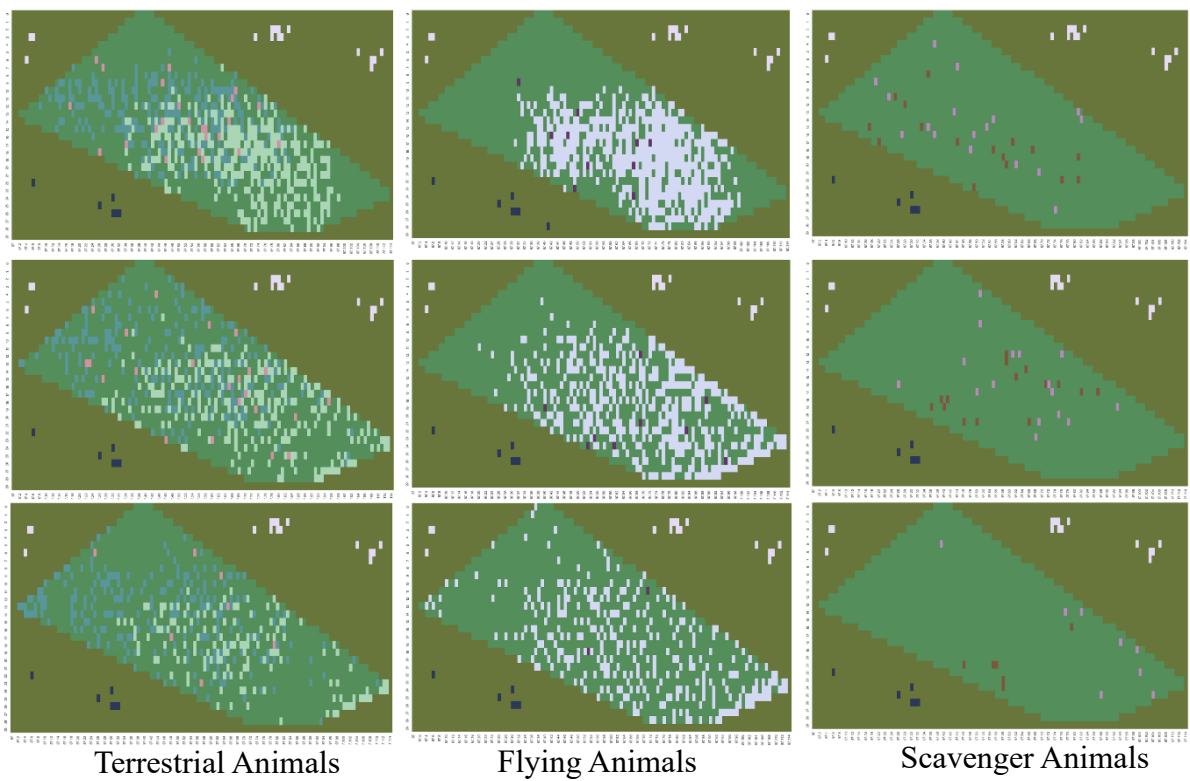

    Terrestrial Animals        Flying Animals        Scavenger Animals

**Figure 7: Changes in the distribution of the seven species over time**

The columns from left to right in Figure 7 are three species of terrestrial animals, two species of flying animals and two species of scavenger animals, respectively. The rows from top to bottom correspond to 1st, 5th and 20th years after the beginning of the ABM simulation.

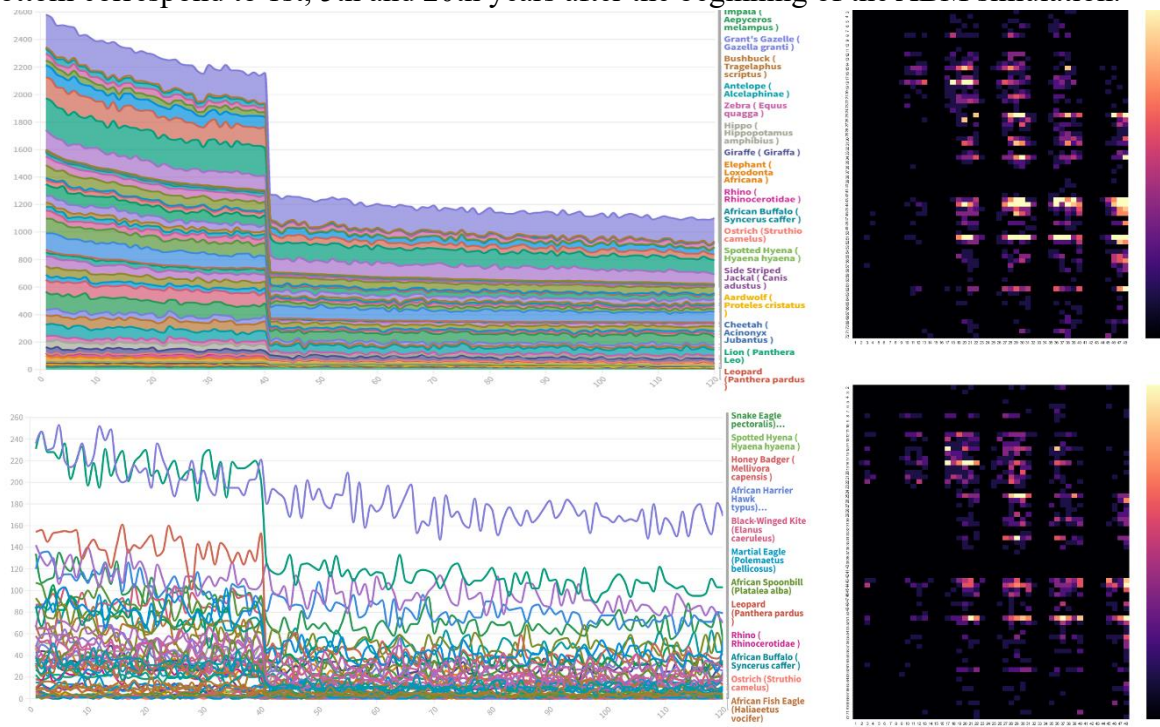

**Figure 8. Changes in the Amount of Animal Agents**

The **two figures in the left column** show the change in the quantity of animal agents **over the 12 years** since the beginning of the ABM simulation, with the **horizontal coordinates in months and the vertical coordinates in population density ratios** of animal agents.

The two heatmaps in the right column show the population density ratios for each of the **71 animal agents in the 48 parcels at the two time points**, and the **brightness of the color blocks** in the heatmaps is a response to the **magnitude of the population density ratios**. The upper heatmap indicates the distribution of animal agents in the parcel at the beginning of the 1st year of the ABM simulation. The lower heatmap indicates the situation at the 5th year of the simulation.

As can be seen from the stacked and line plots on the left, Since the age parameters of the animal agents initialized at the beginning of the simulation are **juvenile, and immature animal agents are not yet able to start reproducing**, resulting in the number of animal agents born during the period of time at the beginning of the simulation is not enough to compensate for the number of deaths. And most animal agents do **not live longer than four years**, so at the time point around the **40th month**, there are several species of animal agents with huge populations and relatively near life spans that die massively around this time point. And the stacking figure visualizes the total population of animal agents, thus the stacking figure shows a steep downward trend. **In the part of sensitivity analysis,** we changed the initial parameters of ABM to avoid rapid population decline around the 40th month.

After a rapid population decline, the total population of animal agents was maintained at a relatively level for a considerable period afterwards. This reflects the **stability and self-sustainability** of the ABM-simulated ecosystem of the preserve before the intervention of humans.

# 5 Human Impact on Agent-based Model (ABM) for Maasai Mara

## 5.1 Simulating the Maasai Mara Preserve with the Intervention of Humans

To model human impacts on the Maasai Mara Reserve using the ABM, we defined 10 types of human agents. These 10 types of human agents are defined according to their status. They are: Short Term Tourists, Long Term Tourists, Maasai Hunters, International Poachers, Preserve Rangers (Kenya), Zoologists, Maasai Tribesmen, Citizens (Kenya), Pastoralists (Tanzania) and Anthropologists. Each human agent is defined by its identity with different attributes and behaviors, which constitute the manner in which human agents interact with the ecosystem, including animal agents.

The pseudo-code for the different types of humans are as follows:

| *Short Term Tourists* | *Long Term Tourists* |
| --- | --- |
| Attributes:<br>- probability_grass_disappears = 0.025<br>- probability_water_contaminated = 0.02<br>- preferred_location = 'roads'<br>- birth_location = 'airport'<br><br>Behavior:<br>- for each walk:<br>   - if in peak season (June to September):<br>     - double number of travellers<br>   - if grass disappears (based on probability):<br>     - decrease health of all animals in range for 2 months<br>   - if water is contaminated (based on probability):<br>     - decrease health of all animals drinking in range for 1 month | Attributes:<br>- probability_grass_disappears = 0.05<br>- probability_water_contaminated = 0.04<br>- preferred_location = 'roads'<br>- birth_location = 'main gate'<br><br>Behavior:<br>- for each walk:<br>   - if grass disappears (based on probability):<br>     - decrease health of all animals in range for 1 year<br>   - if water is contaminated (based on probability):<br>     - decrease health of all animals drinking in range for 1 month |

Both short-term and long-term tourists are people who are attracted to the Maasai Mara Preserve and the animals within it. Both short term tourists and long term tourists are defined as having some extent of negative impact on the environment when acting as human agents. And in turn have an impact on animal agents. Short-term tourists stay in the reserve for a short period of time and their numbers are linearly correlated with the number of animals, doubling during the peak tourist season from June to September. Longer-term visitors stay in the Preserve for longer periods of time and also have a negative impact on the environment.

| *Maasai Hunters:* | *International Poachers* |
| --- | --- |
| Attributes:<br>- preferred_location = 'water'<br>- birth_location = 'north-east to Masai tribe'<br><br>Behavior:<br>- for each hunt:<br>   - prioritize hunting for rhino, lion, and pangolin<br>   - capture only one animal at a time | Attributes:<br>- preferred_location = 'water'<br>- birth_location = 'north-west by base'<br><br>Behavior:<br>- for each hunt:<br>   - prioritize hunting for rhino, lion, and pangolin<br>   - if a rhino is caught, stop hunting for the month<br>   - if another animal is caught, continue hunting for the month |

Both Maasai hunters and international poachers hunt animals in the preserve. However, Maasai hunters hunt some animals in a restricted manner according to tribal customs[9], while international poachers hunt a variety of animals frequently.

| Maasai Tribesmen | Citizens (Kenya) |
| --- | --- |
| Attributes: | Attributes: |
| - annual_economic_trade_index = 0 | - annual_tourism_development_index = 0 |
| - preferred_location = ['roads', 'Maasai tribe'] | - location = 'town' |
| - birth_location = 'Maasai tribe to the north-east' | |
| | Behavior: |
| Behavior: | - for each visit from a tourist: |
| - for each encounter with a tourist or zoologist: |    - update annual_tourism_development_index |
|    - update annual_economic_trade_index | |

Maasai tribesmen are human agents living in the Maasai Mara Preserve who barter or trade money by selling tribal artifacts, including Maasai beaded jewelry, to human agents, such as tourists[10]. And the citizens are human agents living outside of the preserve who mainly serve the tourism industry in the Maasai Mara Preserve and whose income is closely linked to the local economic development.

| Zoologists | Preserve Rangers (Kenya) |
| --- | --- |
| Attributes: | Attributes: |
| - annual_zoological_research_index = 0 | - preferred_location = 'grass' |
| - preference = ['water', 'grass'] | - birth_location = 'town to the north-east' |
| - birth_location = 'airfield' | |
| Behavior: | Behavior: |
| - for each walk: | - for each patrol: |
|    - randomly choose a preference (water or grass) |    - capture only one poacher at a time |
|    - update annual_zoological_research_index | |

Zoologists are human agents who research wild animals in the preserve, and because of their professionalism, there is little damage to the ecosystem[11], and the number of wildlife species they encounter correlates with local scientific research. And the preserve rangers are human agents who patrol the preserve and are tasked with catching international poachers in preserves.

| Pastoralists (Tanzania) | Anthropologists |
| --- | --- |
| Attributes: | Attributes: |
| - insecurity_index = 0 | - annual_anthropological_research_index = 0 |
| - preferred_grazing_time = ['grass', 'home orientation'] | - preferred_location = 'Maasai tribe' |
| - birth_location = 'Tanzanian village to the south-west' | - birth_location = 'airport' |
| | |
| Behavior: | Behavior: |
| - for each grazing period (one year): | - for each encounter with Maasai tribesmen or Tanzanian pastoralists: |
|    - randomly choose a preferred grazing time (grass or home orientation) |    - update annual_anthropological_research_index |
|    - update insecurity_index based on encounters with predators | |

Pastoralists from neighboring Tanzania lead their flocks around the preserve boundaries, but there is a risk of the flock being lost in preserve[12]. Anthropologists interact with pastoralists and Maasai tribesmen to conduct anthropological studies and enhance the local research[13].

## 5.2 Ecosystem Dynamic with Human Intervention

With the addition of human agents to the ABM model, the **distribution of animal agents** at the 1st, 5th and 20th years is shown in the three figures in the leftmost column. The three figures in the second column on the left then show the changes in the **distribution of human agents** at the three corresponding year points. The **stacked and line figures** in the middle column then show how the population of animals changed over the 12 years since the beginning of the simulation after the intervention of humans. The **two heat maps** in the rightmost column show the magnitude of the population density ratios of the various animal agents in each block at the two time points.

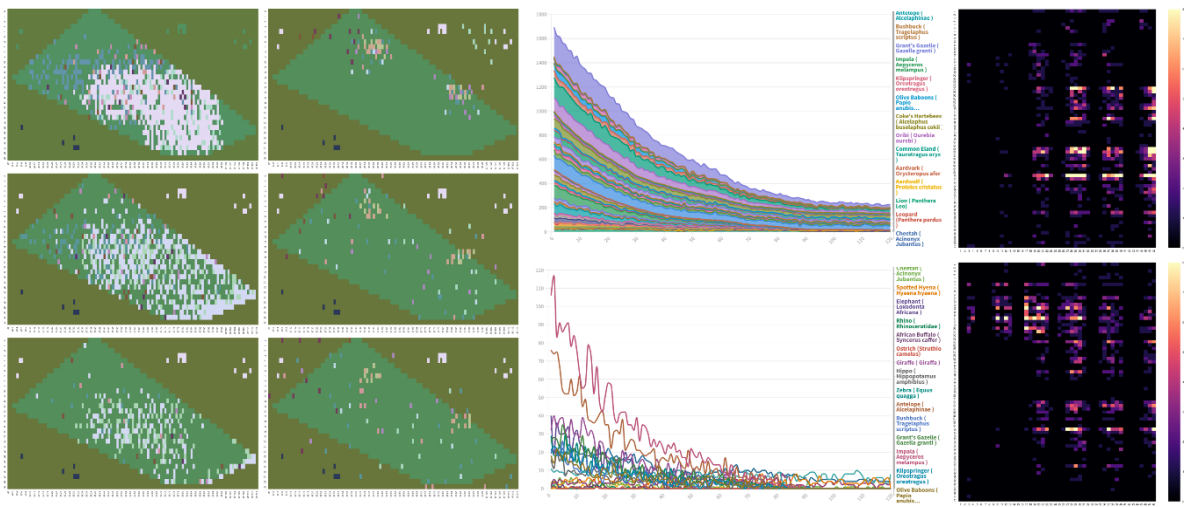

**Figure 9. Changes in the Distribution of Animal Agents and Human Agents**

As seen by Figure 9, when human agents are considered, the number of animal agents of almost all species in the Masai Mara Preserve has been declining continuously for 12 years after the beginning of the simulation. At the end of the simulation, the total number of animal agents in the reserve dropped to about one-fifth of the value at the beginning of the simulation.

At the same time, there have been some changes in the amount and distribution of human agents. Most of the international poachers in the preserve **have been apprehended by preserve rangers**, and almost all of the remaining international poachers are located on the edge of the preserve and it's outside. As the population density ratio of animal agents has decreased, **the number of short-term visitors** has also decreased significantly. So, the tourism **income of local citizens and Maasai tribesmen** has also been decreasing. The **scientific output of zoologists** has also decreased with the decline of animal agents. In the other hand, the distribution and status of anthropologists and Tanzanian pastoralists did not change significantly.

In summary, without policies to limit and manage human activities, animal populations in the Maasai Mara Preserve will decline dramatically, i.e., the ecology is severely damaged. This will reduce the income of the citizens and the Maasai tribesmen and the scientific output of zoologists. This results in an **economic decline** in the region, i.e., **a shrinkage of tourism, and a decline in research output.**

# 6 Development Evaluation Model for the Maasai Mara Preserve

## 6.1 Evaluation by Preserve Development Index

To evaluate the status of the Maasai Mara Preserve and the surrounding area we Preserve Development Index, Ω and the corresponding indicator evaluation model.

Preserve Development Index, Ω is a composite of Environmental Indicator, α, Scientific Research Indicator, β, Economic Indicator, γ and Safety Indicator, δ, which are derived from a combination of TOPSIS and Entropy Weight Method. These four indicators are derived from their corresponding factors by TOPSIS. There are 12 factors used to derive α, β, γ, and δ, respectively, which are derived directly or indirectly by **Monte-Carlo sampling** of all types of agents and parcels.

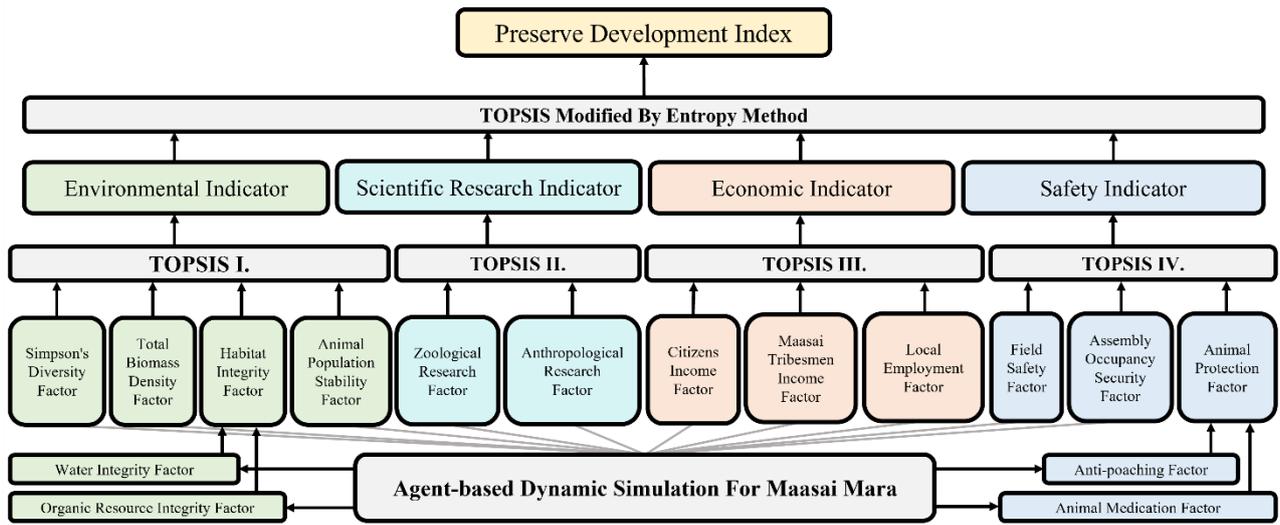

**Figure 10. Hierarchy Diagram of the Evaluation Model**

The Habitat Integrity Factor, $\theta$, is averaged from the Organic Resource Integrity Factor, $\rho$, and the Water Integrity Factor, $\psi$. where $\rho$ and $\psi$ depend on the relative density ratios of organic and organic resources, respectively. Similarly, the Animal Protection Factor, $\upsilon$, is averaged from the Anti-poaching Factor, $\phi$, and the Animal Medication Factor, $x$.

In this case, the Simpson's Diversity Factor, $\zeta$ is given by the following equation:

$$\xi = \frac{\sum n^*(n^* - 1)}{N^*(N^* - 1)} \quad (12)$$

Where $n^*$ is equal to the total number of a particular species, $N^*$ is equal to the total number of water resource and organism resource.

**Monte Carlo sampling** is a statistical method for estimating unknown values or metrics using random sampling. In ABM, Monte Carlo sampling can be used to estimate metrics of agents by randomly sampling from the distribution of agent parameters and behavior.

The basic idea is to run a simulation of the agent-based model multiple times, with different random seeds and random perturbations to the parameters of the agents. Each simulation run generates a sample of agent behaviors and outcomes, which can be used to estimate the factors.

## 6.2 Entropy Weight Method and TOPSIS Method

The entropy weight method is a multi-criteria decision-making method that uses the concept of entropy to calculate the weights of the criteria. The formula for the entropy weight method is as follows:

$$X = \begin{bmatrix} x_{11} & x_{12} & \cdots & x_{1m} \\ x_{21} & x_{22} & \cdots & x_{2m} \\ \vdots & \vdots & \ddots & \vdots \\ x_{n1} & x_{n2} & \cdots & x_{nm} \end{bmatrix} \xrightarrow{\text{Normalization}} z_{ij} = x_{ij} / \sqrt{\sum_{i=1}^{n} x_{ij}^2} \quad (13)$$

Normalization of the normalization matrix of the composition of evaluation indicators and evaluation objects, Obtain the matrix Z. Then calculate the probability matrix P, where each element $P_{ij}$ in P is calculated by the following formula:

$$p_{ij} = \frac{\tilde{z}_{ij}}{\sum_{i=1}^{n} \tilde{z}_{ij}} \quad (14)$$

For the *i*-th indicator, its information entropy is calculated by the formula:

$$e_j = -\frac{1}{\ln n} \sum_{i=1}^{n} p_{ij} \ln(p_{ij}) \quad (j = 1, 2, \cdots, m) \quad (15)$$

By normalizing, it is able to obtain the entropy weight of each indicator:

$$W_j = d_j / \sum_{j=1}^{m} d_j \quad (j = 1, 2, \cdots, m) \quad (16)$$

The Technique for Order of Preference by Similarity to Ideal Solution (TOPSIS) method is a multi-criteria decision-making method that involves finding the ideal and anti-ideal solutions for a set of alternatives. The formula for the TOPSIS method is as follows.

Normalize the decision matrix *X* in Formula (12) to obtain the normalized matrix *Z*:

$$Z = \begin{bmatrix} z_{11} & z_{12} & \cdots & z_{1m} \\ z_{21} & z_{22} & \cdots & z_{2m} \\ \vdots & \vdots & \ddots & \vdots \\ z_{n1} & z_{n2} & \cdots & z_{nm} \end{bmatrix} \quad (17)$$

Find the ideal and anti-ideal solutions, $Z^+$ and $Z^-$, respectively, using the formulas:

$$Z^+ = (Z_1^+, Z_2^+, \cdots, Z_m^+)$$
$$= (\max\{z_{11}, z_{21}, \cdots, z_{n1}\}, \max\{z_{12}, z_{22}, \cdots, z_{n2}\}, \cdots, \max\{z_{1m}, z_{2m}, \cdots, z_{nm}\}) \quad (18)$$

$$Z^- = (Z_1^-, Z_2^-, \cdots, Z_m^-)$$
$$= (\min\{z_{11}, z_{21}, \cdots, z_{n1}\}, \min\{z_{12}, z_{22}, \cdots, z_{n2}\}, \cdots, \min\{z_{1m}, z_{2m}, \cdots, z_{nm}\}) \quad (19)$$

Calculate the distance between each alternative and the ideal and anti-ideal solutions, $D_i^+$ and $D_i^-$, respectively, using the formulas:

$$D_i^+ = \sqrt{\sum_{j=1}^{m} (Z_j^+ - z_{ij})^2} \quad (20)$$

$$D_i^- = \sqrt{\sum_{j=1}^{m} (Z_j^- - z_{ij})^2} \quad (21)$$

Calculate the relative closeness to the ideal solution for each alternative using the formula:

$$S_i = \frac{D_i^-}{D_i^+ + D_i^-} \quad (22)$$

# 7 Specification of Policies and Management Strategies

To enable the Maasai Mara Preserve to develop in a holistic and ecological manner. We have specified 21 policies and management strategies. They are related to **environment, research, economy, security and diplomacy.** Also, some policies and strategies are related to multiple aspects, i.e., they have multiple attributes. At the same time, we classify policy and strategies into **Cluster I and Cluster II** according to their effects on the environment or on the economy, respectively; Cluster III represents a **combination of some of the previous two clusters.**

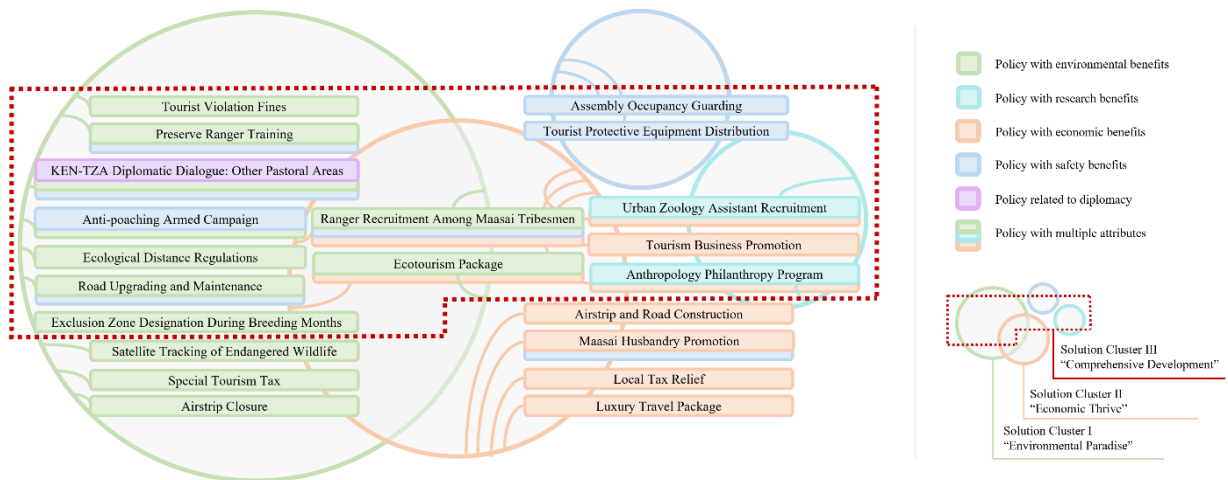

**Figure 11. Policies and Management Strategies with Solutions Clusters They Formed**

**Policy on visitor fines:** *Visitors who violate park rules will be fined 2,000 to 50,000 Kenyan Shilling by park rangers for polluting water and trampling grass. The purpose of this policy is to deter harmful activities and fund park operations.*

**Preserve Ranger Training:** *Rangers will receive professional training to detect and prevent poaching, protect wildlife and grasslands. The training will cover park regulations, wildlife ecology, first aid, communication, conflict resolution, and law enforcement. The purpose of this policy is to enhance ranger effectiveness in managing and protecting the protected area.*

**Ranger Recruitment Among Maasai Tribesmen**: *This management strategy aims to recruit and train qualified Maasai tribesmen as forest rangers. The training will cover forest conservation, wildlife management, monitoring and reporting, and conflict resolution. The goal is to increase local involvement and employment opportunities in forest protection and management.*

**Urban Zoology Assistant Recruitment:** *This policy aims to provide training to qualified individuals from local Kenyan towns to work as zoologist assistants. The training will cover animal behavior, field research techniques, and data analysis. The goal is to increase employment opportunities and enhance the capacity of the zoology sector in Kenya.*

**KEN-TZA Diplomatic Dialogue: Other Pastoral Areas:** *This policy aims to engage in diplomatic dialogue with Tanzanian pastoralists to encourage opening new grazing areas instead of grazing in animal reserves. The goal is to reduce conflicts with wildlife, protect natural habitats, and maintain positive relationships with neighboring countries.*

**Anti-poaching Armed Campaign:** *This management strategy requires clearing known poacher camps every five years to deter poaching, protect wildlife, and natural resources*.

**Tourism Business Promotion:** *This policy provides tax incentives and subsidies to ecotourism*

businesses to promote sustainable and responsible tourism practices and provide economic benefits to local communities.

**Special Tourism Tax:** *This policy increases taxes on tourism businesses to generate additional government revenue for conservation and public services.*

**Ecotourism Package:** *This policy requires eco-tourism companies to educate tourists on environmental protection during tours, to promote sustainable and responsible tourism practices and increase awareness of the importance of protecting natural resources.*

**Ecological Distance Regulations:** *This management strategy prohibits visitors from approaching resource-rich and dense animal areas to protect natural habitats and wildlife, and ensure visitor safety.*

**Exclusion Zone Designation During Breeding Months:** *This management strategy designates core conservation areas where visitors are prohibited from approaching animals during the breeding season to protect them and their young, and minimize human disturbance.*

**Anthropology Philanthropy Program**: *This policy provides project grants to Maasai folklore researchers through the Anthropology Philanthropy Program. The aim is to support the documentation and preservation of Maasai folklore, customs, and traditions, and promote research and cultural exchange.*

**Assembly Occupancy Guarding:** *This management strategy strengthens security at the airport and reserve gates to prevent animal and poacher incursions and protect visitors.*

**Tourist Protective Equipment Distribution:** *This management strategy provides protective gear to visitors and teaches them how to use it to prevent attacks from large wild animals, promoting visitor safety and reducing human-wildlife conflicts.*

**Road Upgrading and Maintenance:** *This policy requires regular maintenance and upgrading of roads to ensure safe and accessible transport for visitors, rangers, and staff, and minimize negative environmental impacts.*

**Satellite Tracking of Endangered Wildlife:** *This management strategy uses satellite tracking to monitor endangered animal populations for conservation efforts, species protection, and anti-poaching efforts.*

**Airstrip Closure:** *This management strategy closes the main entrance (airstrips) to the reserve to reduce visitor numbers and environmental impacts, and promote conservation efforts.*

**Airstrip and Road Construction:** *The government of Kenya will invest in building new airports and roads to support increased tourism while minimizing their impact on animal habitats and the environment.*

**Maasai Husbandry Promotion:** *Establish designated livestock areas within the reserve boundaries, encourage Maasai communities to raise livestock in these areas, provide resources and support to develop sustainable livestock operations.*

**Local Tax Relief :** *The policy aims to reduce taxes for businesses directly supporting the tourism industry in towns near Masai Mara to encourage investment and growth in the local tourism sector.*

**Luxury Travel Package:** *The management strategy allows for the construction of high-end, environmentally sustainable camps in protected areas for long-term, high-spending tourists to increase revenue from Kenya's tourism industry.*

Based on the results of the **ABM simulations**, we measured the changes in the Maasai Mara Preserve's indicators for each of the factors when no strategies were implemented through the evaluation system. As can be seen from Figure 12, most of the factors produced different levels of decreases in their values as the ABM simulation proceeded. The economically related factors such as citizens income factor and employment factor show relatively small decreases, while the ecologically related factors such as Animal Population Stability factor show the largest decreases. However, **almost all of the factors are decreasing in the long term**, which shows that the economic development of the preserve is at the expense of the ecology when **no policy is implemented**, and that this economic development mode is **unsustainable.**

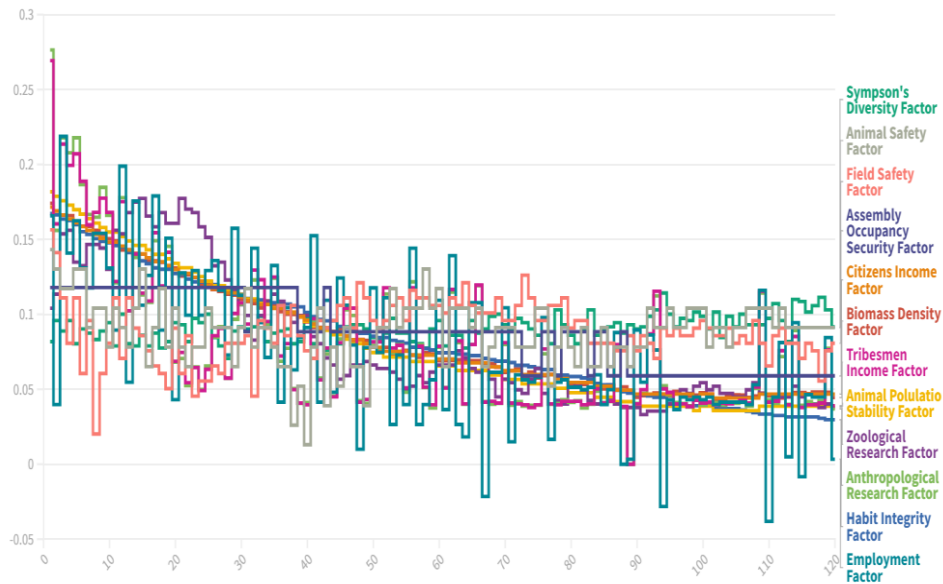

**Figure 12.Changes in factors before the implementation of policies and management strategies.**

For each of the 4 indicators composed of the 12 factors, it can be seen that **all the indicators have decreased**. The safety indicator shows the smallest decrease, and the environmental indicator shows the largest decrease.

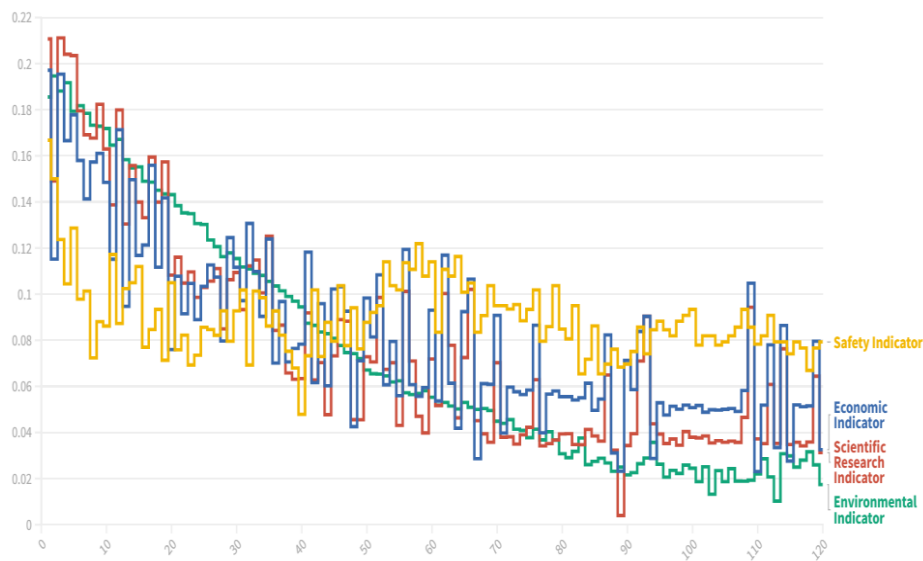

**Figure 13.Changes in indicators before the implementation of policies and strategies.**

Finally, for the **Preserve Development Index (PDI)**, which reflects the **overall development status of Maasai Mara Preserve**, the value of the PDI shows a fluctuating downward trend over time. After 120 years of simulation, the value of PDI decreased to about 0.189, which indicates that the future development of the protected area is **not optimistic in the absence of appropriate policy and management strategy.**

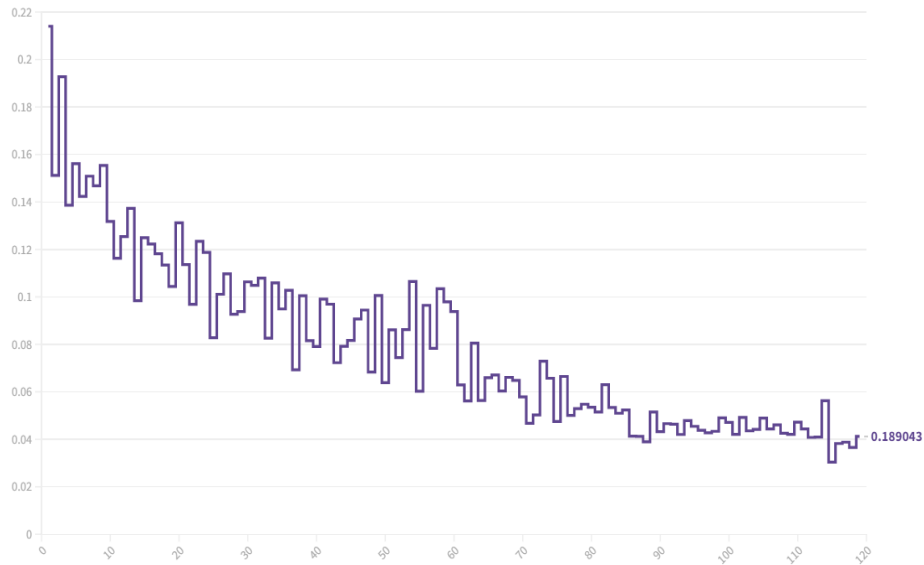

**Figure 14. Changes in the PDI before the implementation of policies and strategies.**

The above figures are without any measures. The three rows of figures presented below are the policies and management strategies for implementing Cluster I, II, and III, respectively. It is clear that the implementation of Cluster III will yield the best results.

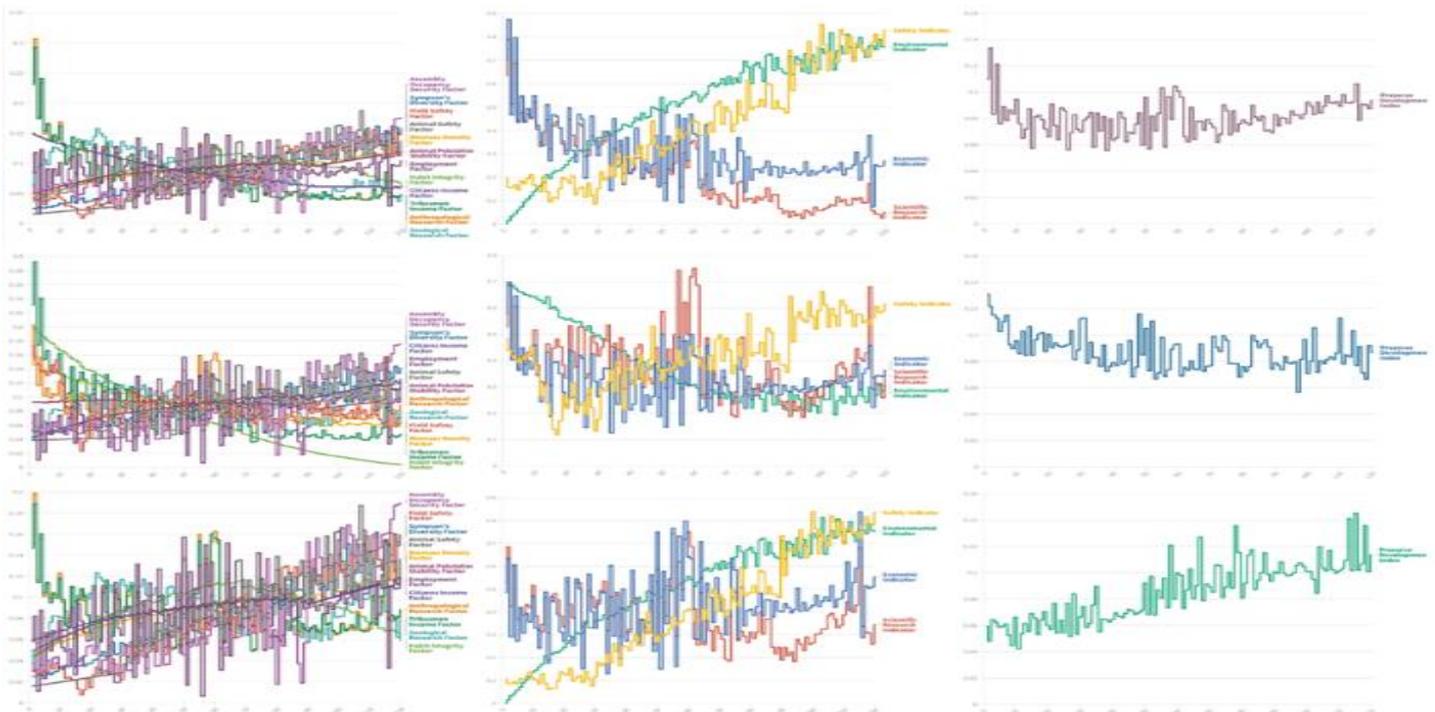

**Figure 15. Changes in indicators after the implementation of policies and management strategies.**

# 8 Sensitivity Analysis

By adjusting the initial parameters i.e., the population density ratios at the beginning of the simulation to the ABM before human intervention, we obtain a stable trend similar to that of the ecosystem before human intervention in the above section. This reflects the strong robustness of this ABM, which ensures the self-sustainability of the ecosystem even after the initial parameter changes.

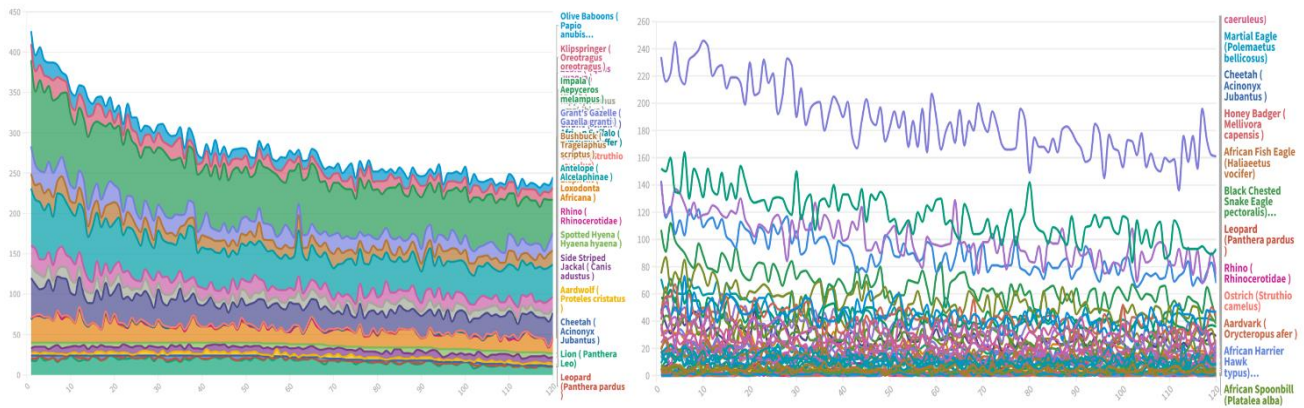

**Figure 16. Population Density Ratio of Animal Agents after Adjusting Initial Parameters.**

In conclusion, the sensitivity analysis of agent-based model show that it can provide valuable insights into the robustness and reliability of the model and can identify the key factors that influence model outcomes. **ABM can also play a curial role in a wide range of applications, including ecology, epidemiology, and social sciences.** By assessing the sensitivity of the model to changes in the input parameters, researchers can improve their understanding of how the model works and make more informed policy decisions. The sensitivity analysis can help identify the most influential factors, which may differ for animal and human agents. The insights from such an analysis can guide the development of more accurate and realistic models, which can in turn inform the design of interventions that improve the health and wellbeing of both animal and human populations.

# 9 Model Evaluation and Further Discussion

## 9.1 Strengths

- Highly interpretable: The **behavior and attributes of specific agents** give **ABM** a highly interpretable degree. Unlike "black box models" such as **neural networks** and **time series**.
- Flexibility: ABM allows for modeling of **complex systems** with non-linear interactions, feedback loops, and emergent behavior. It is highly flexible and can be used to represent a wide variety of systems, ranging from social systems to ecological systems to financial systems.
- Realistic representation of behavior: ABM can **simulate realistic behavior** of individual agents, which helps to understand the system-level behavior. The model can be used to explore different scenarios, assess the impact of policy decisions, and predict future behavior.
- Bottom-up approach: ABM takes a **bottom-up approach**, which means that it starts with modeling the behavior of individual agents and then aggregates their behavior to understand the system-level behavior. This approach allows for a more realistic representation of complex systems, as it can capture the heterogeneity and diversity of individual agents.
- Integration with other models: ABM can be **integrated with other modeling techniques**, such as system dynamics and network analysis, to create a more comprehensive model of a system.

## 9.2 Weaknesses

- High computational requirements: ABM can be computationally intensive, especially when modeling large systems or when the model has a large number of agents. This can result in long simulation times, which can make it difficult to explore different scenarios and conduct sensitivity analyses.
- Difficulty in parameterization: ABM requires the specification of many parameters to define the behavior of the agents. These parameters can be difficult to estimate from empirical data and can introduce uncertainty into the model.
- Lack of standard process: There is currently no standard methodology for ABM, which can make it difficult to compare and replicate results across studies.

## 9.3 Directions of Improvement

- Development of standard methodologies: Developing a standard methodology for ABM can help to ensure that results are comparable and reproducible across studies.
- Integration with empirical data: Integrating empirical data into ABMs can improve their accuracy and relevance for real-world applications. For example, data on individual behavior, social networks, or environmental conditions can be used to inform the development and calibration of the model.
- Validation and verification: Validating and verifying ABM using empirical data can help to ensure that the model is realistic and accurate in representing the system.

# References

[1] *Ambassador Godec's remarks for World Wildlife Day: Speech: Kenya* (2016) *Archive - U.S. Agency for International Development*. Available at: https://2012-2017.usaid.gov/kenya/speeches/ambassador-godec%E2%80%99s-remarks-world-wildlife-day (Accessed: February 17, 2023).

[2] *The Wildlife Conservation and Management Act, 2013.* Republic of Kenya, Kenya Gazette Supplement No. 181 (Acts No. 47), 2013.

[3] Whitlock, M. C. (1996). *The red queen beats the jack-of-all-trades: the limitations on the evolution of phenotypic plasticity and niche breadth.* The American Naturalist, 148, S65-S77.

[4] Haken, H. (2006). *Information and self-organization: A macroscopic approach to complex systems.* Springer Science & Business Media.

[5] Nagy, K. A., Girard, I. A., & Brown, T. K. (1999). Energetics of free-ranging mammals, reptiles, and birds. *Annual review of nutrition,* 19(1), 247-277.

[6] Shorrocks, B., & Bates, W. (2015). *The biology of African savannahs*. Oxford University Press, USA.

[7] Camberlin, P., Moron, V., Okoola, R., Philippon, N., & Gitau, W. (2009). Components of rainy seasons' variability in Equatorial East Africa: onset, cessation, rainfall frequency and intensity. *Theoretical and applied climatology*, 98, 237-249.

[8] Lamprey, H. F. (1963). Ecological separation of the large mammal species in the Tarangire Game Reserve, Tanganyika 1. *African Journal of Ecology*, 1(1), 63-92.

[9] Tarayia, G. N. (2004). The legal perspectives of the Maasai culture, customs, and traditions. *Ariz. J. Int'l & Comp. L.*, 21, 183.

[10] Snyder, K. A., & Sulle, E. B. (2011). Tourism in Maasai communities: a chance to improve livelihoods?. *Journal of Sustainable Tourism*, 19(8), 935-951.

[11] Hunter, R. G. (1984). Managerial professionalism in state fish and wildlife agencies: a survey of duties, attitudes, and needs. *Fisheries*, 9(5), 2-7.

[12] Bhola, N., Ogutu, J. O., Piepho, H. P., Said, M. Y., Reid, R. S., Hobbs, N. T., & Olff, H. (2012). Comparative changes in density and demography of large herbivores in the Masai Mara Reserve and its surrounding human-dominated pastoral ranches in Kenya. *Biodiversity and Conservation*, 21, 1509-1530.

[13] Salazar, N. B. (2010). Towards an anthropology of cultural mobilities. *Crossings: Journal of Migration & Culture*, 1(1), 53-68.